\documentclass[letterpaper,twocolumn,10pt]{article}
\usepackage{usenix-2020-09}

\usepackage[ruled, lined, commentsnumbered, longend]{algorithm2e}
\usepackage{xcolor}
\usepackage{url}
\usepackage{amsmath}
\usepackage{multicol}
\usepackage{listings}
\usepackage{graphicx}
\usepackage{xurl}
\usepackage{tabularx}
\usepackage{comment}
\usepackage{wspr}
\usepackage{algpseudocode}
\usepackage{booktabs}
\usepackage{fancyhdr}
\usepackage{multirow}
\usepackage{ragged2e}
\usepackage{csquotes}
\usepackage{subcaption}
\usepackage{tikz}

\usepackage{enumitem}

\usepackage{xspace}

\newcommand{\totalcombinedVFCDataset}{54,858\xspace}
\newcommand{\totalVFCTrain}{9,143\xspace}
\newcommand{\totalNonVFCTrain}{45,715\xspace}
\newcommand{\totalVFCTrainRepos}{2,658\xspace}

\newcommand{\totalcombinedVFCTypeDataset}{7,847\xspace}

\newcommand{\totalcombinedGHSADataset}{138,529\xspace}

\newcommand{\medianGHSACommitsBetweenVersions}{15\xspace}
\newcommand{\maxGHSACommitsBetweenVersions}{11,015\xspace}

\newcommand{\totalGHSATrainingAdvisories}{2,138\xspace}

\newcommand{\totalGHSATrainingCommits}{138,529\xspace}

\newcommand{\meanGHSATrainingCommits}{15\xspace}

\newcommand{\gtNormalizedCommitRankGHSA}{0.67\xspace}

\newcommand{\nvdInitialPullCVEs}{193,250\xspace}
\newcommand{\nvdGitHubPatchLinks}{8,951\xspace}

\newcommand{\osvGitHubPatchLinks}{3,633\xspace}

\newcommand{\vulasGitHubPatchLinks}{1,282\xspace}

\newcommand{\vfcTypeTotal}{7,847\xspace}
\newcommand{\vfcMissingTypeTotal}{1,296\xspace}

\newcommand{\vfcFinderTotalTopOneAcc}{80.0\%\xspace}
\newcommand{\vfcFinderTotalTopTwoAcc}{89.5\%\xspace}
\newcommand{\vfcFinderTotalTopThreeAcc}{93.2\%\xspace}
\newcommand{\vfcFinderTotalTopFiveAcc}{96.6\%\xspace}

\newcommand{\vfcFinderQoneTopOneAcc}{95.9\%\xspace}

\newcommand{\vfcFinderQtwoTopOneAcc}{90.9\%\xspace}

\newcommand{\vfcFinderQthreeTopOneAcc}{85.8\%\xspace}

\newcommand{\vcMatchonVFCFinderTopOne}{44.0\%\xspace}
\newcommand{\vcMatchonVFCFinderTopFive}{70.1\%\xspace}
\newcommand{\vcMatchonVCMatchTopOne}{89.6\%\xspace}

\newcommand{\vfcIdentificationFOne}{89.3\%\xspace}
\newcommand{\vfcIdentificationAcc}{94.4\%\xspace}

\newcommand{\vfcTypeTopOne}{80.1\%\xspace}
\newcommand{\vfcTypeTopTwo}{88.8\%\xspace}

\newcommand{\vfcTypeTopFive}{98.6\%\xspace}

\newcommand{\missingCleanGHSA}{3,343\xspace}

\newcommand{\totalReviewedGHSA}{334\xspace}

\newcommand{\initialMissingGitHubRepo}{2,129 (34.6\%)\xspace}

\newcommand{\foundMissingGitHubRepo}{1,092\xspace}

\newcommand{\stillMissingGitHubRepo}{1,037 (16.8\%)\xspace}

\newcommand{\potentialRemainMissingGH}{5,122\xspace}

\newcommand{\stillNotFixed}{1,537 (24.9\%)\xspace}

\newcommand{\missingTagsRepo}{239 (3.88\%)\xspace}

\newcommand{\missingPreviousTag}{3\xspace}

\newcommand{\totalGHSAMissed}{12 (3.59\%)\xspace}
\newcommand{\totalGHSACouldNotFind}{25 (7.78\%)\xspace}
\newcommand{\totalGHSAFound}{296 (88.6\%)\xspace}

\newcommand{\totalGHSAFoundAndMissed}{308\xspace}

\newcommand{\totalGHSATopOne}{81.2\%\xspace}
\newcommand{\totalGHSATopTwo}{88.9\%\xspace}
\newcommand{\totalGHSATopThree}{93.5\%\xspace}
\newcommand{\totalGHSATopFive}{96.1\%\xspace}

\usepackage{amsthm}
\theoremstyle{definition}

\usepackage{makecell}

\newcolumntype{C}[1]{>{\centering\arraybackslash}p{#1}}

\newlist{steps}{enumerate}{1}
\setlist[steps, 1]{label = RQ \arabic*:}

\definecolor{codegreen}{rgb}{0,0.6,0}
\definecolor{codegray}{rgb}{0.5,0.5,0.5}
\definecolor{codepurple}{rgb}{0.58,0,0.82}
\definecolor{backcolor}{rgb}{0.95,0.95,0.95}
\definecolor{darkgreen}{RGB}{0,102,0}
\definecolor{darkorange}{RGB}{255, 102, 0}
\definecolor{light-red}{RGB}{222, 98, 98}
\definecolor{light-green}{RGB}{182, 224, 197}
\lstdefinestyle{mystyle} {
  basicstyle=\ttfamily\scriptsize
  frame=single,
  backgroundcolor=\color{backcolor},
  showspaces=false,
  showtabs=false,
  showstringspaces=false,
  breakatwhitespace=false,
  breaklines=true,
  numbers=left,
  numberstyle=\scriptsize,
  numbersep=5pt,
  tabsize=2,
  captionpos=b,
  commentstyle=\bfseries\color{gray},
  keywordstyle=\bfseries\color{red},
  basicstyle=\linespread{1.1}\ttfamily\footnotesize,
  moredelim=[il][\textcolor{lightgray}]{\$\$},
  moredelim=[is][\textcolor{lightgray}]{\%\%}{\%\%}
}
\lstdefinelanguage{diff}{
    basicstyle=\ttfamily\scriptsize,
    morecomment=[f][\color{darkgreen}]{+\ },
    morecomment=[f][\color{red}]{-\ },
}

\begin{document}

\date{}

\title{\Large \bf VFCFinder: Seamlessly Pairing Security Advisories and Patches}



\author{
{\rm Trevor Dunlap}\\
North Carolina State University\\
tdunlap@ncsu.edu
\and
{\rm Elizabeth Lin}\\
North Carolina State University\\
etlin@ncsu.edu
\and
{\rm William Enck}\\
North Carolina State University\\
whenck@ncsu.edu
\and
{\rm Bradley Reaves}\\
North Carolina State University\\
bgreaves@ncsu.edu
} 

\maketitle

\begin{abstract}

Security advisories are the primary channel of communication for discovered vulnerabilities in open-source software, but they often lack crucial information.
Specifically, 63\% of vulnerability database reports are missing their patch links, also referred to as vulnerability fixing commits (VFCs).
This paper introduces VFCFinder, a tool that generates the top-five ranked set of VFCs for a given security advisory using Natural Language Programming Language (NL-PL) models.
VFCFinder yields a \vfcFinderTotalTopFiveAcc recall for finding the correct VFC within the Top-5 commits, and an \vfcFinderTotalTopOneAcc recall for the Top-1 ranked commit.
VFCFinder generalizes to nine different programming languages and outperforms state-of-the-art approaches by 36 percentage points in terms of Top-1 recall. 
As a practical contribution, we used VFCFinder to backfill over 300 missing VFCs in the GitHub Security Advisory (GHSA) database.
All of the VFCs were accepted and merged into the GHSA database.
In addition to demonstrating a practical pairing of security advisories to VFCs, our general open-source implementation will allow vulnerability database maintainers to drastically improve data quality, supporting efforts to secure the software supply chain.

\end{abstract}


\section{Introduction}
\label{sec:intro}

Security advisories help users identify vulnerabilities, apply necessary fixes, and facilitate informed decision-making regarding components within software.
The United States and the European Union have emphasized the need for high-quality advisories to address software dependency vulnerabilities effectively~\cite{supplychain_usa, nis_uk}. 
Nevertheless, many existing security advisories lack crucial information~\cite{dong2019towards}.

\begin{figure}[t]
    \centering
    \includegraphics[width=3.3in]{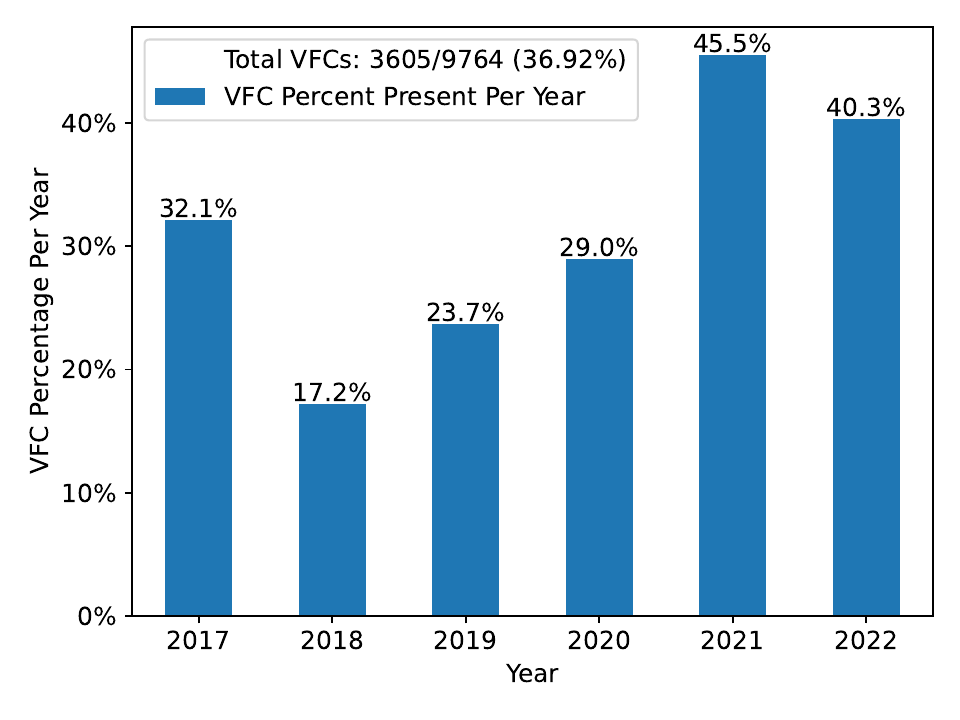}
    \vspace{-1em}
    \caption{
    63.1\% of GHSA security advisoires are missing their patch link based on a snapshot taken through 2022.
    }
    \label{fig:ghsa_vfc_count_20221123}
\end{figure}

Vulnerability fixing commits (VFCs) are a valuable but often missing part of security advisories.
VFCs help practitioners mitigate vulnerabilities by enhancing software composition analysis tools~\cite{pashchenko2020vuln4real, ponta2020detection} 
and enabling patch presence verification~\cite{zhang2018precise, sun2022verjava, wang2023graphspd}, as well as  new state-of-the-art techniques such as enabling few-shot bug repair~\cite{ma2017vurle, xia2023keep}. 
While the security community frequently focuses on identifying new vulnerabilities in code, less attention is given to identifying fixes for vulnerabilities~\cite{hommersom2021mapping,tan2021locating,wang2022vcmatch}.
This disparity is also reflected in practice. 
GitHub and Sonatype use human curators to enhance vulnerability databases~\cite{github_validation, snyk_validation}; however, the volume of security advisories exceeds the available workforce, leading to 63\% of advisories without patch links; see Figure~\ref{fig:ghsa_vfc_count_20221123}.

Prior work established several variations for matching security advisories to VFCs. 
Initial approaches include extracting the vulnerability ID from commit messages~\cite{jimenez2018engineering} or following reference links in advisories~\cite{li_large-scale_2017, xu2022tracking}. 
However, poorly documented security commit messages~\cite{reis2023security} and incomplete security advisories~\cite{dong2019towards} limit the effectiveness of these techniques.
In response to these limitations, machine learning approaches have shown promise by transitioning the task into a ranking problem. 
For instance, FixFinder~\cite{hommersom2021mapping} ranks commits using 23 features and a logistic regression model, achieving a Top-1 recall of 65.1\% and Top-5 recall of 77.7\% on a single Java dataset~\cite{ponta_manually-curated_2019}.
PatchScout~\cite{tan2021locating} uses 22 features and RankNet~\cite{burges2005learning} to attain a Top-1 recall of 69.5\% and Top-5 recall of 85.4\% across various C/C++ projects and a single Java project. 
VCMatch~\cite{wang2022vcmatch}, and its GUI-based implementation Patchmatch~\cite{shen2023patchmatch}, extends PatchScout using 100 features and three machine learning models to achieve the highest reported Top-1 recall of 88.9\% and Top-5 recall of 95.33\% across 10 OSS projects.

\paragraph{Existing Limitations} Despite the reported performance metrics, several key factors limit the application of the current state-of-the-art~\cite{hommersom2021mapping, tan2021locating,wang2022vcmatch} in practice. 
 
\emph{(1)~Lack of Representative Training Data:} 
We performed a preliminary study on OSS projects with fixes in GHSA and found that 41.2\% of the projects do not include any contributing guidelines.\footnote{Evaluating if a CONTRIBUTING.md file exists in a repository.}
Without guidelines, contributors submit poor quality commits messages~\cite{reis2023security}.
However, VCMatch (the top-performing prior work) evaluated rigorously maintained projects with restrictive contributing guidelines. 
For example, FFmpeg's contribution policy mandates that a reference to an issue on the bug tracker is insufficient.
Contributors must also include a summary of the bug in the commit message.\footnote{\url{https://ffmpeg.org/developer.html\#Contributing}}
This dataset is not representative of the broader software supply chain.

\emph{(2)~Non-contiguous Data Sampling:} 
PatchScout and VCMatch build their training and evaluation datasets by randomly selecting project commits without considering commit ordering or relation.
This random sampling can have unintended consequences.
For example, the time difference between the commit and the associated CVE file date in VCMatch can become emphasized by the model as a discriminating feature and could lead to overestimation of recall.
In contrast, a contiguous sampling approach does not introduce this issue.

\emph{(3)~Model Complexity and Risk of Overfitting:}
VCMatch incorporates 100 features, complicating its interpretability and heightening the risk of model overfitting.
Prior machine learning research~\cite{geman1992neural, belkin2019reconciling} shows more features lead to a higher variance and tend to overfit noisy patterns in the training data, resulting in poor accuracy on new examples.

\paragraph{Our System}
In this paper, we propose VFCFinder, a novel approach for helping an analyst match a given security advisory to its VFC.
Our key approach takes the window of commits between the fixed version and the prior version and uses a combination of five intuitive features to produce a ranked set of five potential VFCs for a given advisory.
These features are:
(1)~the likelihood a commit fixed a vulnerability, 
(2)~the type of vulnerability fixed, 
(3)~the similarity between the commit message and the advisory details, 
(4)~where the commit appeared in the window, and 
(5)~any direct indicators in the commit message (i.e., CVE/GHSA-ID).

The first two features, VFC fix probability and VFC vulnerability type, are generated by fine-tuning the CodeBERT NL-PL model~\cite{feng_codebert_2020}.
The semantic similarity between commits and advisory details is generated from sentence embeddings using a pre-trained language model. 
The final two features, commit location and CVE/GHSA-ID in the message, are statically generated.
Finally, these features are fed into an XGBoost model for ranking.
We then introduce a contiguous sampling technique that divides the training and testing sets between fixed and prior versions, simulating the approach a human would take to identify a VFC.

\paragraph{Evaluation and Measurement}
We evaluate VFCFinder in two ways.
First, we construct a representative dataset consisting of the set of all security advisories from the GHSA database with a known patch link: thousands of projects spanning nine programming languages.
VFCFinder identifies the correct VFC for a given security advisory \vfcFinderTotalTopFiveAcc of the time within the Top-5 ranked commits and \vfcFinderTotalTopOneAcc within the Top-1 ranked commit.
For projects with \medianGHSACommitsBetweenVersions or fewer commits between version releases,
VFCFinder identifies the VFC for a given security advisory with a Top-1 recall of \vfcFinderQtwoTopOneAcc.
In contrast, running VCMatch on our dataset resulted in a Top-1 recall of \vcMatchonVFCFinderTopOne and a Top-5 recall of \vcMatchonVFCFinderTopFive. 

Second, we deploy VFCFinder on over 300 randomly selected GHSA advisories without patch links to demonstrate that VFCFinder generalizes beyond our training and testing data.
VFCFinder found the missing patch link with a Top-5 recall of \totalGHSATopFive and a Top-1 recall of \totalGHSATopOne.

In summary, we make the following key contributions.

\begin{itemize}
    \item \textit{We propose a security advisory-to-VFC matching approach that generalizes to nine programming languages and thousands of open-source projects. 
    }
    In contrast to prior work, which uses 100 features~\cite{wang2022vcmatch}, our approach only uses five. 
    By using a smaller set of features, we reduce the amount of variance in the resulting model, making the model generalize. 
    Specifically, VCMatch~\cite{wang2022vcmatch} has a 36 percentage point lower Top-1 recall than VFCFinder when evaluated on VFCFinder's dataset, which spans thousands of projects and nine languages.
    Whereas VFCFinder has a similar performance to VCMatch when tested on the VCMatch dataset (not included in VFCFinder's training).

    \item \textit{We propose a new evaluation standard for security advisory-to-VFC matching tools.}
        Prior work~\cite{tan2021locating, wang2022vcmatch} uses a non-contiguous sampling approach for VFC ranking, which overestimates their recall in practice.

    \item \emph{We deployed VFCFinder to backfill over 300 security advisories in the GitHub Security Advisory database.}
        GitHub's security team confirmed all of our submitted VFCs and integrated them into the GHSA database.
\end{itemize}

\paragraph{Availability}
A version of VFCFInder is available at \url{https://github.com/s3c2/vfcfinder}.
\section{Background}
\label{sec:background}

\begin{figure}[t]
    \centering
    \includegraphics[width=3.1in]{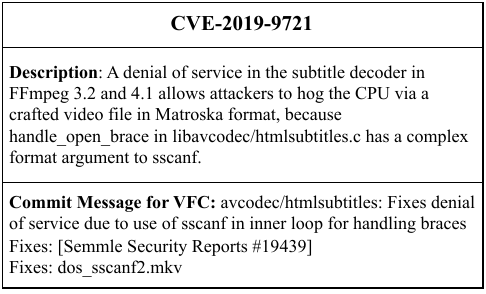}
    \caption{An informative commit message for a VFC within FFmpeg for CVE-2019-9721.}
    \label{fig:CVEFFmpeg}
\end{figure}

Security advisories play a crucial role in the overall health of the software supply chain. 
The U.S. Department of Homeland Security and the Cybersecurity and Infrastructure Security Agency sponsor the MITRE corporation in maintaining the Common Vulnerability Enumeration (CVE) Program~\cite{cve_process}. 
This program lets researchers and vendors report identified vulnerabilities in software. 
The reported vulnerabilities are then publicly given tracking numbers called CVE IDs by the CVE Assignment Team and CVE Numbering Authorities. 
The CVE IDs are then openly placed on the CVE list~\cite{cve_list}.

Various downstream databases, such as the National Vulnerability Database (NVD)~\cite{nvd}, GHSA~\cite{ghsa}, and Google's Open Source Vulnerabilities (OSV)~\cite{osv_ref}, then access the CVE list to enhance reports further. 
For instance, NVD adds fields for numerical scoring of vulnerability severity, such as the Common Vulnerability Scoring System score. 
The GHSA database converts free-form text fields from the CVE into machine-readable fields, such as fixed versions. 
The CVE project emphasizes the importance of including accurate and adequate information in the initial vulnerability description, highlighting that omitting key details can create complications downstream. 
One significant detail within the CVE is the reference link to the patch link. 

\begin{figure}[t]
    \centering
    \includegraphics[width=3.1in]{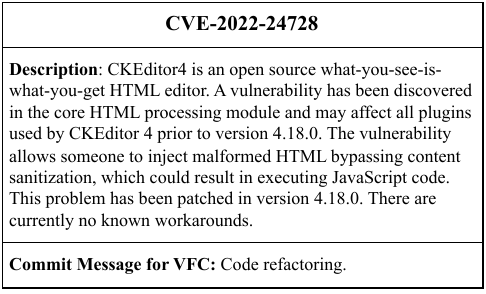}
    \caption{A misleading commit message for the VFC within CKEditor4 for CVE-2022-24728.} 
    \label{fig:CVECKEditor}
\end{figure}

\paragraph{Motivating Examples}
Figures~\ref{fig:CVEFFmpeg} and \ref{fig:CVECKEditor} present two scenarios of documenting vulnerabilities and VFCs.
Figure~\ref{fig:CVEFFmpeg} shows CVE-2019-9721, from the VCMatch dataset, which addressed a denial of service within FFmpeg. 
The advisory description is clear. 
The commit message addresses the vulnerability and is similar to the advisory, making it straightforward to pair the advisory to the patch.
VCMatch and VFCFinder rank the commit as the top choice for the VFC.

In contrast, for a cross-site scripting vulnerability in the CKEditor4 project (Figure~\ref{fig:CVECKEditor}), the CVE description is informative, but the commit message for the patch link contains little useful information.  
Had the patch link not been included in the original CVE reference links, it would have been nearly impossible for a human to identify. 
There are 78 commits between the fixed version (4.18.0) and the prior version (4.17.2).
VCMatch ranked the corresponding commit as 38th, whereas VFCFinder ranked the commit as third.

\paragraph{Multiple VFCs and Multiple Versions}
We anecdotally observed that some security advisories reference multiple VFCs. 
To understand this relationship, we performed a preliminary study and found that 96\% of GHSA security advisories with a fix (see Section~\ref{sec:ghsa_data_collection} for data collection process) only list a single VFC.
The remaining 4\% of advisories offer more than one VFC. 
Of those, 39\% are for patches in multiple versions. 
For instance, the project \emph{parse-server} from GHSA-2m6g-crv8-p3c6 gives two patch links that correspond to two patched versions (4.10.14 and 5.2.5). 
In fact, the backport patch link for 4.10.14 needed more changes than the patch for version 5.2.5, demonstrating that it is important to find all VFCs.
Therefore, identifying the patches for each reported version is valuable to aid practitioners.

\textbf{Commit Window for VFCs:}
We hypothesized that VFCs generally appear between the reported fixed version and the prior version.
To test this hypothesis, we performed a second preliminary study that examined all GHSA security advisories with a VFC and found that around 65\% of VFCs appear in this range. 
An additional 29\% of commits are backported from the VFC listed in the security advisory, indicating that there exists a VFC in the hypothesized range. 
Therefore, 94\% of the examined security advisories had the VFC in the expected location. 
The remaining VFCs not appearing in the anticipated location may be due to unreliable version data~\cite{nguyen2013reliability, anwar2021cleaning}.
VFCFinder leverages this intuition for its approach.

How the commit window is determined matters.
PatchScout's optional branch analysis is the closest to using a commit window: it considers all commits for an entire branch.
However, version releases within GitHub are based on git tags, not branches~\cite{git_releases}.
For example, \emph{parse-server} maintains branches 4.x.x and 5.x.x, each having many minor releases and CVEs.
Therefore, the commit window should be based on git tags and not branches.

\begin{figure*}[t]
    \centering
    \includegraphics[width=7in]{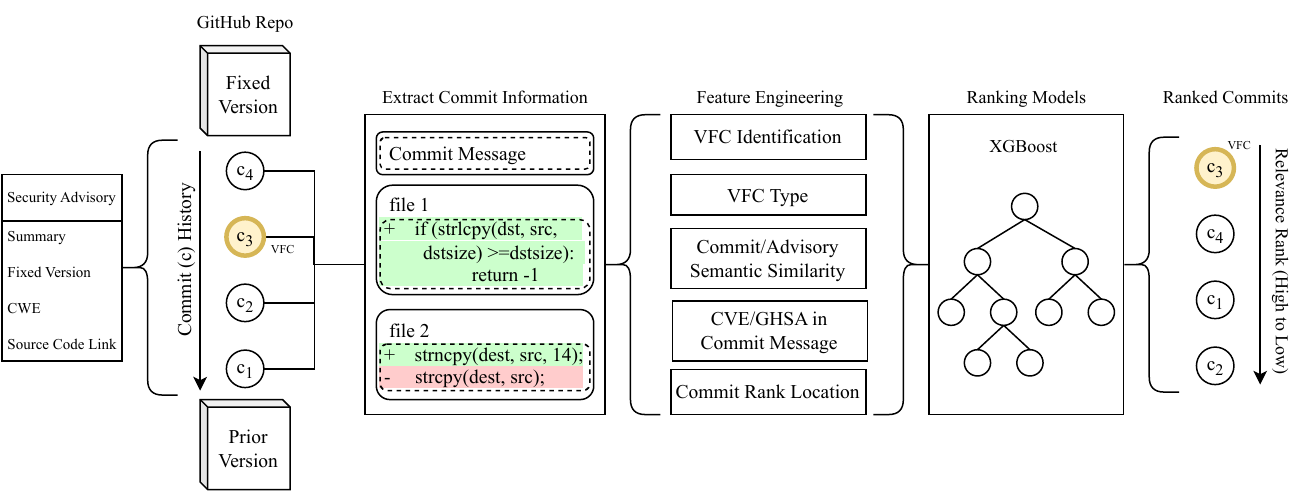}
    \caption{
        The architecture of VFCFinder illustrates ranking commits based on their relevance to fixing a given security advisory.
    }
    \label{figures:Overview_Figure_VFC}
\end{figure*}

\section{The Design of VFCFinder}
\label{sec:model}

Figure~\ref{figures:Overview_Figure_VFC} illustrates the architecture of VFCFinder.
The primary goal of VFCFinder is to match security advisories to VFCs. 
Initially, VFCFinder consumes a security advisory extracting information regarding the fixed and prior version commit windows.  
Then, considering each commit within the window, the commit message and git diff are extracted. 
Leveraging CodeBERT~\cite{feng_codebert_2020}, VFCFinder generates the first feature, predicting the likelihood that a commit fixed a vulnerability. 
We fine-tune the CodeBERT model using data from NVD~\cite{nvd}, OSV~\cite{osv_ref}, and VulasDB~\cite{ponta_manually-curated_2019}. 
In addition to VFC identification, VFCFinder uses CodeBERT for vulnerability type classification for each VFC, explicitly focusing on the OWASP Top 10. 
The third feature is a commit-to-advisory semantic similarity score using SentenceTransformers. 
The final two features are a CVE or GHSA identifier in commit messages and the commit rank location. 
These features are fed into a single XGBoost model to create the final ranking of commits relevant to fixing security advisory.

\subsection{Extracting Advisory Information}
\label{sec:extracting_info}
VFCFinder uses the OSV format~\cite{osv_format} for security advisories, which provides the following key-value data : (1) a detailed vulnerability summary, (2) CWE type, (3) source code repository, (4) related CVE/GHSA identifiers, and (5) fixed versions.
VFCFinder additionally identifies the associated VFC for each fixed version. 

\paragraph{Determining Prior Version and Commit Window}
VFCFinder uses git tags, typically used for versioning, to determine the commit window.
Once cloning a project locally, all project tags are retrieved (i.e., via \texttt{git tag}).
The fixed version from the advisory is then matched directly to the tag set.
For the prior version tag, VFCFinder uses the package \emph{packaging},\footnote{\url{https://pypi.org/project/packaging/}} allowing for semantic version sorting of the tags.
The tag immediately preceding each fixed version is selected as the prior tag.
Upon obtaining the fixed and prior tags, the command \texttt{git tag prior\_ver...fixed\_ver} lists all commits within the specified commit window.

\subsection{VFC Identification}
\label{sec:vfc_identification}
The first feature VFCFinder predicts is if a commit resolves a vulnerability, a process based on a fine-tuned CodeBERT model.
CodeBERT~\cite{feng_codebert_2020} is a transformer-based architecture~\cite{vaswani_attention_2017} equipped with bimodal pre-training for natural language (NL) and programming language (PL).
CodeBERT was initially trained on six programming languages paired with function-level documentation.
HuggingFace~\cite{wolf2020transformers} hosts CodeBERT with pre-trained weights, allowing fine-tuning of the model for specific tasks.

\begin{figure}[t]
    \centering
    \includegraphics[width=3.2in]{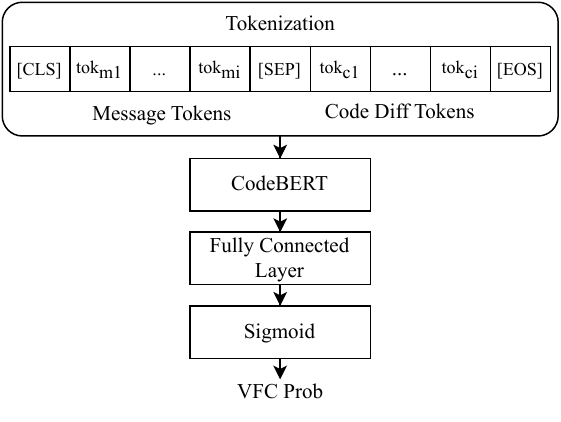}
    \caption{A fine-tuning CodeBERT framework for VFC identification. The fully connected layer of the VFC classification is vector size one. VFC Type identification uses the same framework, but the vector size from the fully connected layer is 10 (i.e., OWASP Top 10) and uses a softmax function.}
    \label{fig:codebert}
\end{figure}

We fine-tune CodeBERT for VFC identification using a custom tuning loop (shown in Figure~\ref{fig:codebert}) and tuning data described in Section~\ref{sec:data_collection}.

\paragraph{Tokenization}
Before fine-tuning, we transcribe the free-form commit message and code into numerical forms through tokenization.
The tokenizer expects two elements from commit data:
(a) the commit message, and (b) the git diff featuring modified, deleted, and added code.
The tokenizer produces a tensor divided into three sections: \emph{input\_ids}, \emph{attention\_mask}, and \emph{token\_type\_ids}.
The \emph{input\_ids} are a blend of the commit message and git diff as follows: \texttt{[CLS] commit\_message [SEP] git\_diff [EOS]}.
Tokens \texttt{[CLS][SEP][EOS]} are special separators; \texttt{[CLS]} signifies the beginning of the segments, \texttt{[SEP]} is a divider between the commit message and raw git diff code, and \texttt{[EOS]} is the end-of-sequence token.
The \emph{attention\_mask} assists the model in identifying \emph{input\_ids} padded tokens, indicating which tokens require attention.
The \emph{token\_type\_ids} designates the start and end of sequences, specifically, the length of the commit message tokens and the git diff.

The tokenizer accepts a maximum token count based on the pre-trained model; for CodeBERT, it is 512 tokens. 
Excess tokens are truncated.
In order to minimize truncation, we section the commit data into smaller chunks, each based on a file with changes, and generate tensors from these chunks. 
This method not only aids in reducing data truncation but also allows us to make predictions and evaluations for individual programming languages separately.

\paragraph{Fine-Tuning}
We implemented a classification fine-tuning loop for the CodeBERT model.
The model includes an embedding layer that maps input tokens to 768-dimensional vectors and 12 encoder layers.
These encoder layers incorporate a self-attention mechanism for focusing on varying parts of the input sequence.
Each encoder's intermediate layer executes a non-linear input transformation, followed by a linear output layer transformation.
The last encoder layer's output is directed to a pooling layer, averaging the hidden states across the input sequence.
This output is then processed through a fully connected layer with an output size of one.
During tuning, we use an unweighted binary cross entropy loss function defined as:
\begin{equation} \label{eq1}
    \begin{split}
        l_{BCE} = & -[y \cdot \log x + (1- y) \cdot \log(1-x)]
    \end{split}
\end{equation}

where $x$ is the input and $y$ is the target.
The logits are passed to a sigmoid activation function, producing the final prediction, a floating value ranging from 0 to 1, indicating the VFC positive class probability.

\paragraph{Aggregating Predictions to Commit Level}
VFCFinder generates predictions on a per-file basis.
This strategy ensures that different programming languages are handled separately during the prediction process.
For instance, when a commit updates Python and C files, CodeBERT does not need to process multiple languages simultaneously.
Therefore, VFCFinder then consolidates file predictions into a total commit prediction.
To do so, VFCFinder calculates the arithmetic mean of the file predictions, resulting in a single value between 0 and 1, where 1 suggests a likely vulnerability resolution.

\subsection{VFC Vulnerability Type}
\label{sec:vfc_type}
The fine-tuning for VFC type identification mimics the VFC identification outlined in Section~\ref{sec:vfc_identification}, differing primarily in the classification tasks.
VFC type is categorized based on the OWASP Top 10 and an additional ``Other'' class that signifies vulnerabilities outside the OWASP Top 10.
Initially, we contemplated predicting VFC type at the CWE level, but since MITRE defines 933 different CWE types and the relatively sparse training data, we decided against it.
Discussion of VFC type data collection and mapping OWASP Top 10 labels to VFC types is in Section~\ref{sec:vfc_type_data_collection}.

\paragraph{Tokenization}
The tokenizer for VFC type is the same for VFC identification, as seen in Section~\ref{sec:vfc_identification}.

\paragraph{Fine-Tuning}
The fine-tuning architecture for VFC type is similar to that of VFC identification.
The primary differences are the output size of the fully connected layer, the loss function, and the activation function.
The VFC type's output size is 10, denoting its deployment for a 10-class classification task.\footnote{As outlined in Section~\ref{sec:vfc_type_data_collection}, the classification size would be 11, but no examples exist for one of the OWASP Top 10 classes.}
We specifically use a weighted cross entropy loss function as defined:
\begin{equation} \label{eq1}
    \begin{split}
        l_{WCE} = & -w_{y} \log \frac{e^{x,y}}{\sum_{c=1}^{10} e^{x,c}} \cdot y
    \end{split}
\end{equation}

where $x$ is the input, $y$ is the target, $w$ is the weight, and $c$ is the number of classes.
A softmax activation function is then used on the fully connected output layer, transforming the results into a probability distribution across the classes.

\paragraph{Aggregating Predictions to Commit Level}
Predicting the VFC type on a per-file basis requires a distinct commit-level aggregation process. 
To determine the VFC type, we use the file prediction that has the maximum probability. 
In specific terms, we use an argmax function to identify and select the OWASP Top 10 type that has the highest probability within a given predictions.
\begin{equation} \label{vfc_type_aggregation}
argmaxf(X) := {x : f(s) \leq f(x), \forall s \in X}
\end{equation}
This procedure ensures the selection of the VFC type with maximum confidence.
The evaluation of VFC type identification is in Section~\ref{sec:eval_vfc_type}.

\subsection{Semantic Similarity}
\label{sec:semantic_similarity}
VFCFinder also incorporates the similarity between the commit message and the original advisory.
For instance, consider the advisory GHSA-fj7c-vg2v-ccrm and its associated VFC:
\begin{displayquote}
    \emph{GHSA-fj7c-vg2v-ccrm description:} ``Buffer leak on incoming WebSocket PONG message(s) in Undertow before 2.0.40 and 2.2.10 can lead to memory exhaustion and allow a denial of service.''
\end{displayquote}
\begin{displayquote}
    \emph{undertow@c7e84a0 VFC commit message:} ``[UNDERTOW-1935] - buffer leak on incoming websocket PONG message''
\end{displayquote}

VFCFinder uses SentenceTransformers~\cite{reimers2019sentence}, an advanced technique for generating embeddings to produce semantic similarity scores between texts.
Specifically, we use the pre-trained all-mpnet-base-v2 model.~\footnote{\url{https://huggingface.co/sentence-transformers/all-mpnet-base-v2}}
VFCFinder then feeds these embeddings into a cosine similarity function to identify semantic correlations from the embeddings.
The output ranges from -1 (indicating opposite meanings) to 1 (denoting identical meanings).
A score of 0 signifies orthogonality or dissimilarity between the two vectors.

Regrettably, not every advisory and VFC commit message is as descriptive as the previous instance.
Take the advisory GHSA-rgp5-m2pq-3fmg and the related VFC as an example:
\begin{displayquote}
    \emph{GHSA-rgp5-m2pq-3fmg description:} ``microweber prior to version 1.2.11 is vulnerable to cross-site scripting''
\end{displayquote}
\begin{displayquote}
    \emph{microweber@f7f5d41 VFC commit message:} ``update''
\end{displayquote}
In the initial example, the cosine similarity score is 0.88, reflecting considerable similarity. 
However, for the second example, despite being the VFC for the advisory, the cosine similarity score is -0.01.

\subsection{Static Features}
\label{sec:static_features}
VFCFinder also incorporates two static features to enhance the classification.
We initially considered other static features, similar to those in prior work~\cite{hommersom2021mapping, tan2021locating, wang2022vcmatch}, however, most demonstrated limited feature importance, leading us to retain the following two prominent static features.

\paragraph{CVE/GHSA Identifier}
In some cases, developers mention the CVE or GHSA identifiers for advisories directly in commit messages. 
Naturally, VFCFinder should encapsulate this information.
The presence of the CVE/GHSA-ID within the commit message is determined using a direct search method. 
This feature is encoded as a binary value, with 1 signifying a match.

\paragraph{Normalized Commit Rank Location}
In our feature engineering, we observed that VFCs often occur towards the commit lifecycle's end, typically before the next version release. 
For instance, the GHSA-prrh-qvhf-x788 advisory resolved a vulnerability across 32 commits, with the VFC (314456d) as the 31st commit, directly preceding the v5.0.2 release.\footnote{\url{https://github.com/PrestaShop/productcomments/compare/v5.0.1...v5.0.2}} 
VFCFinder computes $commit_{rank}/commit_{total}$ for the normalized commit rank location, yielding a location of $31/32=0.97$ for the cited VFC. 
According to our ground truth dataset (Section~\ref{sec:data_collection}), the average normalized commit rank location for VFCs is \gtNormalizedCommitRankGHSA.
This is intuitive for vulnerability patching practices.
As vulnerabilities are discovered, we expect a new software release with the patch provided shortly after an issue is resolved.

\subsection{Ranking Commits}
\label{sec:ranking_models}

\begin{table}[t]
    \centering
    \caption{VFCFinder's ranking model uses five features. 
    }
    \label{tbl:feature_engineering}
    \small
    \begin{tabular}{p{1.0in}p{2.0in}} \toprule
        \multicolumn{1}{c}{\textbf{Features}} & \multicolumn{1}{c}{\textbf{Description}} \\ \midrule
    VFC Probability              & Probability distribution of commit fixing a vulnerability  \\ \midrule
    VFC Type Match$^\ast$               & Boolean match between advisory and VFC Type prediction  \\ \midrule
    Commit/Advisory Similarity   & Similarity score commit message and advisory report \\ \midrule
    CVE/GHSA ID \newline in Commit$^\ast$       & Boolean match if CVE/GHSA ID in commit message  \\ \midrule
    Commit Location              & Normalized commit rank location of a commit in version lifecycle \\ \bottomrule
    \multicolumn{2}{p{3.0in}}{\footnotesize $^\ast$ We describe this as five features, but the XGBoost model uses seven features. We split individual features for CVE and GHSA, and split VFC type into Top-1 and Top-5.}
    \end{tabular}
    \normalsize
\end{table}

The final step in VFCFinder is to use the previously described features, Table~\ref{tbl:feature_engineering}, to rank the commits relevant to the given security advisory. 
VFCFinder uses XGBoost~\cite{chen2016xgboost}, an iterative gradient-boosting algorithm that progressively incorporates decision trees while adjusting observation weights based on previous inaccuracies. 
By combining weak learners and predicting residuals from prior trees, XGBoost uses regularization techniques to optimize performance and mitigate overfitting. 

To initially tune the hyperparameters of the XGBoost model, we used hyperopt~\cite{bergstra2013making}, a Bayesian optimization algorithm. 
The best results were obtained when the learning rate was set to 0.001, and the decision tree depth in the model was restricted to four. 
Furthermore, we set the maximum number of boosting rounds, i.e., the number of decision trees included in the model, to 1,500.

A binary logistic objective was used during training, classifying each commit as related or unrelated to the security advisory fix. 
The model outputs the predicted probabilities for each input to belong to the positive class, which range from 0 (non-match) to 1 (match). 
This process transforms the task into a classification problem. 
These probabilities are then ranked to denote the likelihood of each commit fixing a security advisory. 
Section~\ref{sec:evaluation} elaborates on the model's evaluation.

\section{Data Collection}
\label{sec:data_collection}
Here, we discuss the training and testing datasets for VFCFinder. 
The data collection process is organized into three sets, each corresponding to a unique classification model: VFC identification, VFC type identification, and the final XGBoost ranking process.
Table~\ref{tbl:datasets} provides a summary and the aggregate commit count for each set.

\subsection{Vulnerability Fixing Commits}
\label{sec:vfc_data_collection}
We sourced data from three vulnerability databases: NVD\cite{nvd}, OSV\cite{osv_ref}, and VulasDB~\cite{ponta_manually-curated_2019}. 
NVD, operated by NIST, is a primary vulnerability disclosure platform. 
Google's OSV aggregates data from multiple sources (GHSA, PyPA, RustSec, Global Security Database, and OSS-Fuzz) and primarily targets open-source dependencies. 
VulasDB manually curated vulnerable commits within Java projects and has been used extensively in previous studies~\cite{nguyen2022vulcurator, giang_hermes_2022, sabetta_practical_2018}.

\begin{table}[t]
\centering
\caption{Datasets used for training various aspects of VFCFinder. The dataset size indicates commit count.}
\label{tbl:datasets}
\small
\begin{tabular}{lcc} \toprule
    \multicolumn{1}{c}{\textbf{Dataset}} & \multicolumn{1}{c}{\textbf{Objective}}  & \multicolumn{1}{c}{\textbf{Size}} \\ \midrule
VFC Identification & VFC and Non-VFCs   & \totalcombinedVFCDataset \\
VFC Types & OWASP Top 10 Labeled VFCs  & \totalcombinedVFCTypeDataset \\
GHSA Data & Matching VFCs to advisories   & \totalcombinedGHSADataset \\ \bottomrule
\end{tabular}
\normalsize
\end{table}

Our training data included \totalcombinedVFCDataset commits, with \totalVFCTrain vulnerability-fixing and \totalNonVFCTrain non-vulnerability-fixing commits. 
The commits span nine languages.
We started data collection by downloading all datasets, beginning with the NVD, which contained \nvdInitialPullCVEs CVEs as of August 16, 2022. 
CVEs contain reference links tagged as a \emph{patch}, typically signifying VFCs. 
We focused specifically on GitHub commit links, identifying \nvdGitHubPatchLinks GitHub commit patch links. 
A similar methodology on OSV yielded \osvGitHubPatchLinks GitHub-referencing VFCs, and VulasDB provided \vulasGitHubPatchLinks Java commits linked to open-source dependencies. 
After consolidation and de-duplication, \totalVFCTrain unique VFCs were obtained.

\begin{figure}[t]
    \centering
    \includegraphics[width=3.4in]{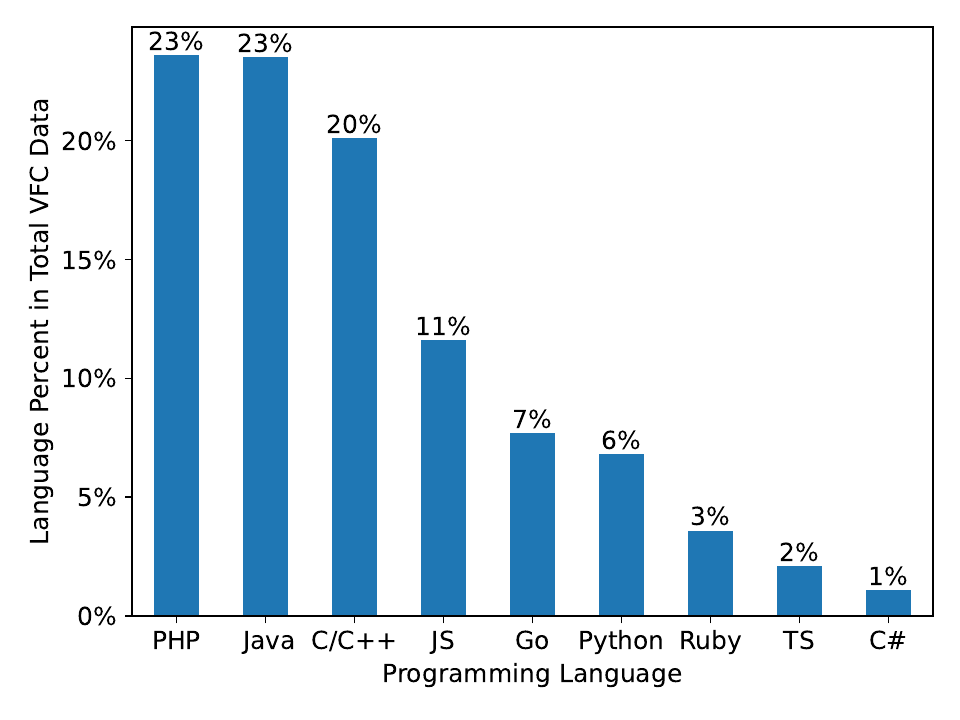}
    \caption{Language breakdown across Vulnerability Fixing Commits within the VFC Identification dataset}
    \label{fig:language-vfc}
\end{figure}

We then cloned each repository containing the commit and confirmed the commits used languages in C/C++, PHP, Java, JavaScript, Python, Go, Ruby, TypeScript, or C\#.
To extract data, we developed an additional tool for parsing commit data, \emph{patchparser}.\footnote{This tool has been publicly released on PyPI: \url{https://pypi.org/project/patchparser/}}
The tool pulls key features from GitHub commits and formats them in a way ideal for ML/AI applications.
Figure~\ref{fig:language-vfc} demonstrates the language distribution for the study, emphasizing PHP and Java.

\paragraph{Collecting Non-VFCs}
Training requires non-VFCs in addition to VFCs. 
Consistent with prior work, we keep a ratio of five non-VFCs to one VFC~\cite{sabetta_practical_2018}. 
For every VFC in a repository, we gather five non-overlapping non-VFCs from the same repository.

To collect non-VFCs, we follow methods established in previous studies~\cite{sabetta_practical_2018, giang_hermes_2022}. 
Firstly, we collate unique repositories from the VFCs, a total of \totalVFCTrainRepos. From these repositories, we compile the commit history. 
We then run a modified version of git-vuln-finder~\cite{git_vuln_finder}, which includes additional keywords from SPI~\cite{zhou_spi_2021}, on the commit history. 
Commits not matched by git-vuln-finder are assumed to be non-VFCs. 
We then check if the commit modified at least one file associated with the study's target languages; if not, the commit is discarded. 
After verifying the commits, we shuffle the non-VFCs and match each VFC with five non-VFCs from the same repository.
We randomly sampled 100 non-VFCs to ensure they weren't related to security fixes. This process yielded \totalNonVFCTrain non-VFCs from \totalVFCTrainRepos unique repositories.

\begin{table}[t]
    \centering
    \caption{VFC OWASP Top 10 Distribution}
    \label{tbl:owasp_eval_data}
    \small
    \begin{tabular}{lc} \toprule
        \multicolumn{1}{c}{\textbf{Category}} & \multicolumn{1}{c}{\textbf{VFC Count}} \\ \midrule
    A01: Broken Access Control & 1,333 \\ 
        A02: Cryptographic Failures & 126 \\ 
        A03: Injection & 2,249 \\
        A04: Insecure Design & 232 \\ 
        A05: Security Misconfiguration & 125 \\
        A06: Vulnerable and Outdated Components & 0 \\
        A07: Identification and Authentication Failures & 322 \\
        A08: Software and Data Integrity Failures & 209 \\
        A09: Security Logging and Monitoring Failures & 30 \\
        A10: Sever-Side Request Forgery (SSRF) & 88 \\
        Other (Weaknesses outside OWASP Top 10) & 3,133 \\\bottomrule
     \end{tabular}
     \normalsize
\end{table}

\subsection{Vulnerability Fixing Commit Types}
\label{sec:vfc_type_data_collection}
Security advisories are associated with a common weakness enumeration (CWE), denoting the type of vulnerability. 
With 933 existing CWE types~\cite{owasp_cwe_map}, predicting a VFC's corresponding vulnerability type is challenging. 
We leverage the CWE to OWASP Top 10 mapping provided by MITRE~\cite{owasp_cwe_map}, simplifying our prediction classes. 
VFCs not falling within the OWASP Top 10 are categorized as ``Other.''

We excluded advisories with multiple CWEs.
Our preliminary analysis found that less than 2\% of the total advisories list multiple CWEs. 
Furthermore, \vfcMissingTypeTotal VFCs did not possess a CWE label, resulting in a dataset of \vfcTypeTotal VFCs with a CWE label. 
Table~\ref{tbl:owasp_eval_data} details the commit distribution per OWASP Top 10 label and the ``Other'' class. 
This imbalance mirrors the real-world vulnerability distribution. 
Notably, "Vulnerable and Outdated Component" was not a classification within the VFC dataset.

\subsection{GHSA Data}
\label{sec:ghsa_data_collection}
GHSA advisories were used to train the final ranking model of VFCFinder due to their guaranteed presence on GitHub and prevalent use for open-source projects.
An advisory was considered for training if it contained a GitHub code repository link and an identified fixed version.
Approximately 42\% of GHSA advisories lack a code repository link, and roughly 28\% remained unpatched, preventing their inclusion in the training process.

Our total dataset consisted of \totalGHSATrainingAdvisories projects and \totalGHSATrainingCommits associated commits.
The average number of commits between fixed and prior versions was \meanGHSATrainingCommits.
Commit labels were determined based on their association with an advisory, with VFCs in the advisory receiving a label of 1 and others labeled as 0.

\paragraph{Contiguous Data Sampling}
The contiguous aspect is to obtain all of the commits in the order in which they appear in the commit lifecycle between the prior and fixed versions. 
As discussed previously, current state-of-the-art~\cite{tan2021locating, wang2022vcmatch} uses a non-contiguous data sampling technique, selecting non-associated commits randomly throughout the project.
FixFinder~\cite{hommersom2021mapping} uses a contiguous sampling approach, but the boundaries are set without respect for the prior and fixed version, creating a selection of commits two years before and one hundred days after the CVE file date for each evaluation.

Additionally, our approach separates training and testing datasets to keep advisories distinct and ensures that commit lifecycles within each set are non-overlapping. 
This strategy ensures the integrity of our training and testing sets, preventing any associated commits from being split between them. 
Further details on training and testing can be found in Section~\ref{sec:evaluation_setup}.

\section{Evaluation}
\label{sec:evaluation}

This section presents the evaluation of VFCFinder on the datasets from Section~\ref{sec:data_collection}. 
We pose four research questions:

\begin{enumerate}[label=\emph{\bf{RQ\arabic*:}},leftmargin=2.5\parindent]
    \item \emph{What is VFCFinder's effectiveness in pairing security advisories and vulnerability fixing commits?}
    This question assesses the full VFCFinder ranking pipeline.
    We benchmark VFCFinder against VCMatch \cite{wang2022vcmatch} on their dataset and across our dataset, representing the software supply chain.

    \item \emph{How well does VFCFinder identify VFCs?} 
    We evaluate VFCFinder against nine programming languages for identifying VFCs. 

    \item \emph{How effective is VFCFinder in determining the VFC type?} 
    Extending past VFC identification, we evaluate how VFCFinder can identify the vulnerability type fixed during the VFC.
    We classify based on the OWASP Top 10 and an ``Other'' class. 
    
    \item \emph{What features are important for matching security advisories to VFCs?} 
    In addition, we provide insight into how the features of matching security advisories to VFCs impact the output of VFCFinder. 
\end{enumerate}

\subsection{Evaluation Setup}
\label{sec:evaluation_setup}
Our evaluation depends on three datasets: GHSA commits (Section~\ref{sec:ghsa_data_collection}, RQ1), VFC/Non-VFC labels (Section~\ref{sec:vfc_data_collection}, RQ2), and VFC types (Section~\ref{sec:vfc_type_data_collection}, RQ3).
These are real-world, up-to-date data from maintained vulnerability databases.

We create a holdout set of 10\% of each dataset, preserving vulnerability type and language imbalances through stratified sampling.
We apply a 5-fold cross-validation for model fine-tuning on the remaining 90\% of data.
Each fold results in a model that we test on the holdout set.
We then average the model probabilities to create the final holdout set prediction.
We confirmed the training/testing and holdout data do not contain any forms of overlap, which would result in data leakage. 
We fine-tune and evaluate models on a machine with an Intel i7-9700k CPU, 32GB RAM, and an NVIDIA RTX 3090 Ti GPU running Ubuntu 20.04.

\subsection{RQ1: Evaluation Results}

\paragraph{Baseline Comparison}
We focus our comparison on VCMatch~\cite{wang2022vcmatch}, as it is the latest and highest reporting metrics for advisory to VFC matching.
The source code and trained models are publicly accessible within Patchmatch~\cite{shen2023patchmatch}, a GUI-based implementation. 
This availability allows for a noise-free, direct comparison. 
Additionally, we omit PatchScout~\cite{tan2021locating} from our comparative analysis because the source code is not publicly available. 
VCMatch replicated PatchScout in their work and already demonstrated a 17-percentage-point performance advantage over PatchScout.
For rigor, we first validated we could replicate VCMatch's results using their original dataset before comparing its performance with our dataset.

\begin{table}[t]
    \centering
    \caption{A Top-N recall comparison of VCMatch~\cite{wang2022vcmatch} vs VFCFinder on VCMatch's dataset (10 OSS projects) and VFCFinder's dataset (2,138 projects). VCMatch's performance on unseen data (VFCFinder data) indicates overfitting, while VFCFinder demonstrates robust performance on new unseen data.}
    \label{tbl:vcmatch_baseline}
    \footnotesize
    \setlength{\tabcolsep}{4pt} 
    \begin{tabular}{l | c c | c c | c c} 
    \toprule
    & \multicolumn{2}{c|}{\textbf{VCMatch}} & \multicolumn{2}{c|}{\textbf{VFCFinder}} & \multicolumn{2}{c}{\makecell{\textbf{Difference}$^*$}} \\ 
    \textbf{Dataset (\# pkgs)} & \textbf{Top-1} & \textbf{Top-5} & \textbf{Top-1} & \textbf{Top-5} & \textbf{Top-1} & \textbf{Top-5} \\ 
    \midrule
    VCMatch (10) & 89.6\% & 94.3\% & 81.9\% & 97.3\% & -7.7 & +3.0 \\
    VFCFinder (2.1k) & 44.0\% & 70.0\% & 80.0\% & 96.6\% & +36.0 & +26.6 \\ 
    \bottomrule
    \multicolumn{7}{p{3in}}{$^*$ VFCFinder performance minus VCMatch performance.}
    \end{tabular}
    \normalsize
    \end{table}

Table~\ref{tbl:vcmatch_baseline} presents the results of VFCFinder compared to VCMatch. 
VFCFinder significantly outperforms VCMatch in Top-1 recall by 36 percentage points (\vfcFinderTotalTopOneAcc vs. \vcMatchonVFCFinderTopOne) on our dataset, demonstrating greater generalizability. 
Although VCMatch shows a marginal 7.68 percentage point increase in Top-1 recall when evaluated on its dataset (\vcMatchonVCMatchTopOne vs. 81.9\%), it suggests overfitting to its specific data. 
Furthermore, VFCFinder excels in Top-5 recall on both datasets, indicating a broader and more consistent ability to identify vulnerabilities correctly. 
These performance metrics in both Top-1 and Top-5 recalls validate VFCFinder's robustness and adaptability in diverse, real-world scenarios for matching security advisories to patch links. 

\begin{figure}[t]
    \centering
    \includegraphics[width=3.3in,clip,trim=0.1in 0.1in 0.1in 0.1in]{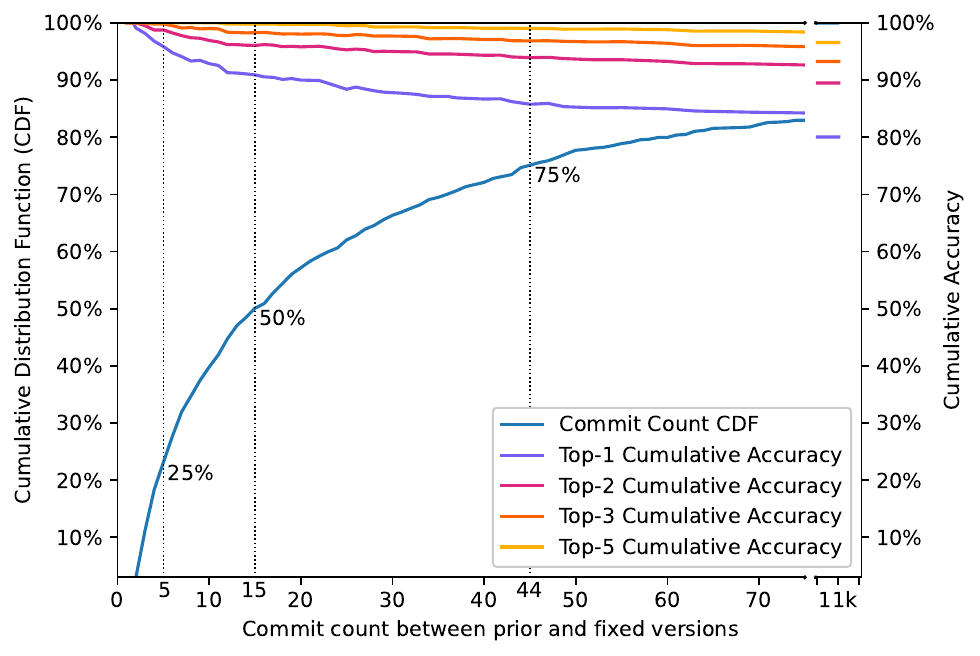}
    \caption{Top-N recall for ranking commits based on the commits between the prior and fixed versions of an advisory. Note, the CDF percentage of the commit count is at 100\% at \maxGHSACommitsBetweenVersions commits (the maximum number of commits seen between versions).}
    \label{fig:final_vfcfinder_results}
\end{figure}

\paragraph{Detailed VFCFinder performance}
Figure~\ref{fig:final_vfcfinder_results} illustrates the performance of VFCFinder.
When searching for the vulnerability fixing commits, each version lifecycle will have a different number of commits. 
We found the median number of commits between versions to be \medianGHSACommitsBetweenVersions.
Focusing on the Top-1 recall, when considering the lower quartile (25\%) of data, the commit count is less than 5; the recall is \vfcFinderQoneTopOneAcc. 
This recall value changes to \vfcFinderQtwoTopOneAcc for the median (50\%, commit count $\leq$ \medianGHSACommitsBetweenVersions) and to \vfcFinderQthreeTopOneAcc for the upper quartile (75\%, commit count $\leq$ 44). 
For the entire dataset, the Top-1 recall is \vfcFinderTotalTopOneAcc. 

For 75\% of the data, the Top-2 to Top-5 accuracies consistently remain above 93\%. 
Upon examining the entire dataset, the Top-2, Top-3, and Top-5 recalls are \vfcFinderTotalTopTwoAcc, \vfcFinderTotalTopThreeAcc, and \vfcFinderTotalTopFiveAcc, respectively. 
However, a slight decrease in Top-N recall and a reduction in commit count at Top-5 occurs.

\paragraph{Accurately Ranked Commits}
Consider GHSA-h47x-2j37-fw5m,\footnote{\url{https://github.com/advisories/GHSA-h47x-2j37-fw5m}} 
an advisory addressing a critical injection vulnerability in the Infinispan project. 
This advisory reports two patched versions and provides two VFCs. 
In the case of version v9.4.17, with 63 commits in the window, VFCFinder correctly ranked the corresponding VFC as first. 
The VFC identification probability was 0.96, and the model accurately classified the vulnerability type. 
Despite the final prediction output of VFCFinder being 0.30, it was significantly higher than the second-rank commit of 0.08.
For the older version, v8.2.12, with just six commits within the window,
VFCFinder also identified the correct VFC as the top-ranked commit, validating patches for both versions.

\paragraph{Misranked Commits}
An instance of an incorrectly ranked commit happened with advisory GHSA-wqv4-9gr3-3qgh related to Jenkins, where VFCFinder ranked the actual VFC seventh among 82 commits.
This version had six additional GHSA-IDs, three of which fell into the same OWASP category, leading to the misranking. 
This highlights the challenge in correctly associating a VFC with its relevant advisory, particularly when multiple vulnerability fixes of a similar type exist between versions.

\vspace{1.0em}
\noindent
\begin{tikzpicture}
    \draw (-10,0) node[draw=black,fill=black!2, thick, rounded corners,inner sep=1ex,text width=0.464\textwidth] {
      {\bf Takeaway: }
      Within five commits, VFCFinder produces a \vfcFinderTotalTopFiveAcc recall for containing the correct VFC within the prior and fixed versions. 
      Over prior work~\cite{wang2022vcmatch}, VFCFinder increases the Top-1 recall by 36\% percentage points when applied to various OSS projects.
      };
\end{tikzpicture}

\subsection{VFCFinder Characterization}
This section assesses the distinct elements of VFCFinder, with a particular emphasis on the fine-tuned CodeBERT models used for VFC identification and VFC type. 
Although a substantial body of research exists regarding VFC identification and type (refer to Section~\ref{sec:relwork}), they predominantly concentrate on individual facets of VFCFinder. 
To our knowledge, a comprehensive analysis of VFC identification spanning nine languages has not been extensively explored. 
Our claim is not to have enhanced VFC identification or VFC type identification; rather, our focus has been to further evaluate CodeBERT's proficiency across these nine languages in VFC identification and VFC type as they are important aspects of VFCFinder. 

\subsubsection{RQ2: VFC Identification}
\label{sec:eval_vfc_indentification}
The VFC identification component of VFCFinder proves effective, achieving an \vfcIdentificationFOne macro F1 score and a \vfcIdentificationAcc accuracy. 
Additionally, the performance generalizes across nine languages. 
We use a base threshold of 0.5 during evaluation to represent a VFC; values below this do not indicate a VFC.
Formally,
\begin{equation}
VFC = 
\begin{cases} 
    0 & \text{if } \frac{1}{n} \sum_{i=1}^{n} x_i < 0.5, \\
    1 & \text{if } \frac{1}{n} \sum_{i=1}^{n} x_i \geq 0.5
\end{cases}
\end{equation}
where $x$ is the output from the sigmoid activation function from the fine-tuned CodeBERT model.

\paragraph{VFC Identification Evaluation} 
Table~\ref{tbl:vfc_language_eval} presents VFC identification results for the dataset outlined in Section~\ref{sec:data_collection}. 
Key metrics include a macro F1 score of 89.3\%, recall of 87.5\%, precision of 91.5\%, accuracy of 94.4\%, and an area under the ROC curve (AUC) of 95.7\%. 
Weighted F1 on the holdout data is 94.2\%. Evaluation against the holdout set yielded 4,452 true negatives, 209 false negatives, 100 false positives, and 704 true positives.

\paragraph{Correctly Classified VFCs}
The true negatives and positives from the model classification within the holdout dataset offer insight into VFCFinder's performance.
The average probability for the 4,452 true negatives is 0.04, with a standard deviation 0.07. 
The true positives average a probability of 0.94, which indicates high model certainty.
VFCFinder also applies to less descriptive commit messages, such as d158413 in ckeditor/ckeditor4, to resolve a cross-site scripting vulnerability with a commit message of ``Code refactoring'' and a probability of 0.84.

\paragraph{Misclassified VFCs}
False negatives refer to true VFCs wrongly classified as non-VFCs, while false positives denote non-VFCs incorrectly classified as VFCs.
False negatives had a mean probability of 0.18, with a standard deviation of 0.16.
For instance, CVE-2017-5553 identified a cross-site scripting (XSS) vulnerability in the b2evolution CMS project, with a single patch link: ce5b36e.\footnote{\url{https://github.com/b2evolution/b2evolution/commit/ce5b36e44b714b18b0bcd34c6db0187b8d13bab8}}
The commit message, \emph{Ignore wrong URLs on markdown plugin}, corresponded to a patch where developers refined an existing regex to accept only URLs beginning with \emph{http://}, \emph{https://}, or \emph{/}.
The model overlooked this subtle regex adjustment and vague commit message, marking it as a false negative.
The probability outputted by VFCFinder was 0.30.
Revising the probability thresholds could make the model identify it as a VFC.

\begin{table}[t]
\centering
\caption{VFC Identification Language Generalization
}
\label{tbl:vfc_language_eval}
\small
\begin{tabular}{lccc} \toprule
      \multicolumn{1}{c}{\textbf{Language}}   & \textbf{Macro Precision} & \textbf{Macro Recall} & \textbf{Macro F1} \\ \midrule
C/C++      & 92.4\%         & 89.3\%      & 90.7\%  \\
Python     & 90.1\%         & 87.6\%      & 88.8\%  \\
TypeScript & 86.1\%         & 86.1\%      & 86.1\%  \\
JavaScript & 89.2\%         & 84.9\%      & 86.9\%  \\
PHP        & 92.9\%         & 88.4\%      & 90.4\%  \\
Java       & 91.0\%         & 84.2\%      & 87.1\%  \\
Ruby       & 93.9\%         & 88.7\%      & 91.0\%  \\
C\#        & 87.5\%         & 98.5\%      & 92.1\%  \\
Go         & 89.4\%         & 85.7\%      & 87.4\%  \\ \midrule
\textbf{Total}      & 91.5\%         & 87.5\%      & 89.3\%  \\ \bottomrule
\end{tabular}
\normalsize
\end{table}

False positives produced a mean probability of 0.72 with a standard deviation of 0.16, implying model uncertainty compared to the mean of 0.93 for true positives.
For instance, a false positive arose from the Ansible package's bug fix for a missing dependency. 
Though the commit message, \emph{defend against bad or missing crypt}, initially suggested a vulnerability fix, code review clarified the issue as a failure due to a missing package.

\paragraph{VFC Language Generalization}
Table~\ref{tbl:vfc_language_eval} presents the performance metrics for the nine languages in our holdout set.
VFCFinder performed well across each programming language.
C\#, with a 92.1\% macro-F1 score, performed best, largely owing to a high recall of 98.5\%, despite its relatively lower precision at 87.5\%.
Interestingly, C\# accounted for the smallest training samples in our dataset.
TypeScript was the least successful, with an 86.1\% macro F1 score.

\vspace{1.0em}
\noindent
\begin{tikzpicture}
    \draw (-10,0) node[draw=black,fill=black!2, thick, rounded corners,inner sep=1ex,text width=0.464\textwidth] {
      {\bf Takeaway: }
      VFCFinder identifies VFCs with an F1 score of 89.3\% and generalizes across nine languages.
      };
\end{tikzpicture}

\begin{table}[t]
\centering
\caption{VFC Type Identification (OWASP-Top 10 + Other Class) Top-N Evaluation}
\label{tbl:vfc_type_eval}
\small
\begin{tabular}{lcccc} \toprule
& \textbf{Precision} & \textbf{Recall}  & \textbf{F1} & \textbf{Accuracy} \\ \midrule
Top-1 & 80.2\% & 80.1\% & 79.7\% & 80.1\% \\
Top-2 & 89.3\% & 88.8\% & 88.5\% & 88.8\% \\
Top-3 & 94.4\% & 94.3\% & 94.1\% & 94.3\% \\
Top-5 & 98.6\% & 98.6\% & 98.6\% & 98.6\% \\ \bottomrule
\end{tabular}
\normalsize
\end{table}

\subsubsection{RQ3: VFC Type Identification}
\label{sec:eval_vfc_type}
Table~\ref{tbl:vfc_type_eval} shows the Top-N metrics for VFC vulnerability type classification, labeled as per the OWASP Top 10.
VFCFinder scores \vfcTypeTopOne, \vfcTypeTopTwo, and \vfcTypeTopFive accuracy for the Top-1, Top-2, and Top-5 labels, respectively.
The metrics are weighted to account for data imbalance.

\begin{figure}[t]
    \centering
    \includegraphics[width=3.0in,clip,trim=0.45in 0.20in 0.85in 0.15in]{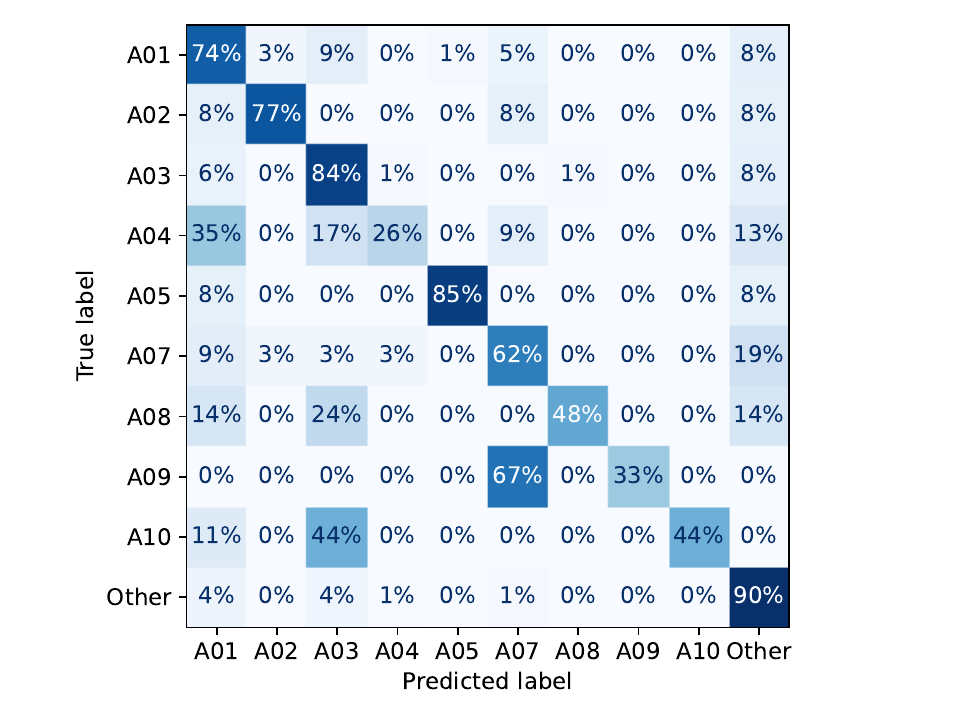}
    \caption{Normalized Confusion Matrix for the VFC Type Top-1 Identification by OWASP Top 10 Categories}
    \label{fig:vfc_type_confusion}
\end{figure}

\paragraph{Correctly Classified VFC Types}
Figure~\ref{fig:vfc_type_confusion} shows the normalized confusion matrix for predicting vulnerability types, labeled by OWASP Top 10 categories as per Table~\ref{tbl:owasp_eval_data}.
The matrix's diagonal indicates the recall for each category; for instance, VFCFinder correctly identified 74\% of A01: Broken Access Control vulnerability types.
The model showed robust performance in the ``Other'' class, correctly detecting 90\% of them for the Top-1 label.
Even with approximately 40\% of the data classified as ``Other,'' VFCFinder can accurately distinguish different OWASP classes.
We anticipate enhanced VFC type classification for other classes with more training data.

\paragraph{Misclassified VFC Types}
Figure~\ref{fig:vfc_type_confusion} additionally shows the misclassification analysis of OWASP Top 10 categories.
For instance, 44\% of A10: SSRF VFCs were predicted as A03: Injection.
Despite initial concerns, these types exhibit notable similarities.
Taking CVE-2022-1723 as an example, an SSRF was fixed in jgraph/drawio before version 18.0.6, mitigating potential local file access by web server attackers.
The commit message "18.0.6 release" fails to specify the patch's purpose.
Upon commit review, a \emph{sanitizeUrl(String url)} function emerged to validate URL parameters, a method similar to A03: Injection patching.
Thus, SSRF and Injection patches may resemble each other regarding code modification.

\vspace{1.0em}
\noindent
\begin{tikzpicture}
    \draw (-10,0) node[draw=black,fill=black!2, thick, rounded corners,inner sep=1ex,text width=0.464\textwidth] {
      {\bf Takeaway: }
      VFCFinder correctly classifies the VFC type with a Top-1 accuracy of \vfcTypeTopOne and a Top-5 accuracy of \vfcTypeTopFive.
      };
\end{tikzpicture}

\subsection{Feature Importance}

\begin{figure}[t]
    \centering
    \includegraphics[width=3in, clip,trim=0.1in 0.1in 0.1in 0.1in]{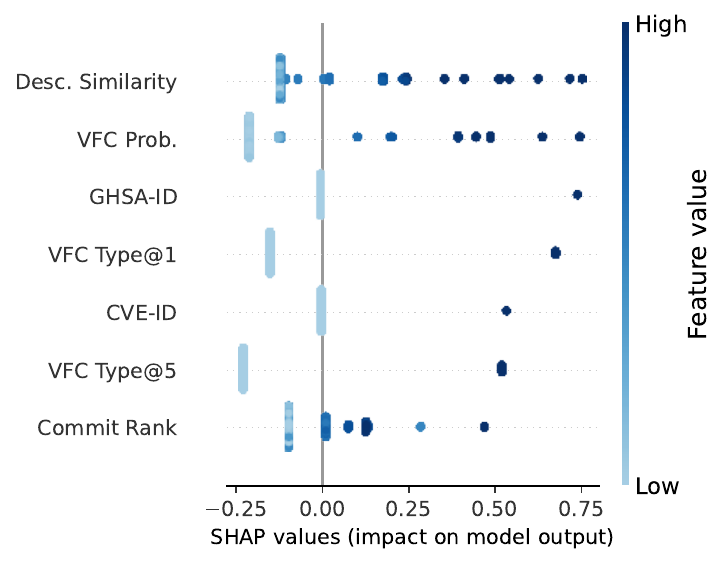}
    \caption{
        The SHAP summary of VFCFinder's features assigns a score to each sample (e.g., Similarity score from -1 to 1). 
    High feature values correspond to high SHAP values, underscoring the importance of all features in VFCFinder's classification. Features are ranked by importance, from highest (Similarity) to lowest (Commit Rank). 
    }
    \label{fig:xgboost_shap_summary}
\end{figure}

In this section, we explore the impact of five specific features outlined in Table~\ref{tbl:feature_engineering} on the performance of VFCFinder. 
Although machine learning models are often seen as black boxes, using SHAP (Shapley Additive Explanations) values~\cite{lundberg2017unified} has enhanced our ability to interpret these models. 
SHAP values serve as metrics for quantifying the importance of individual features.
We approach each combination of features as a distinct power set, subjecting each to training. 
By measuring the marginal contribution of each feature to the model's overall predictive outcome, we assess its relative importance. 
These measured values allow for an understanding of each feature's role in the model.

Figure~\ref{fig:xgboost_shap_summary} visually represents the SHAP values distribution across the complete training data pipeline for VFCFinder. 
The y-axis marks the feature value, with darker colors denoting higher values. 
The x-axis, on the other hand, reflects the aggregated SHAP value. 
A rise in VFC probability correlates with elevated SHAP values, impacting the model's overall predictive probability. 
Conversely, a decrease in VFC probability diminishes the likelihood of a match. 
Features with higher values, such as VFC probability, matching description, and types of vulnerability fixes, are more likely to correspond to an accurate patch link.

\vspace{1.0em}
\noindent
\begin{tikzpicture}
    \draw (-10,0) node[draw=black,fill=black!2, thick, rounded corners,inner sep=1ex,text width=0.464\textwidth] {
      {\bf Takeaway: }
      Each feature for VFCFinder provides a strong contribution to the overall matching of security advisories to VFCs. 
      };
\end{tikzpicture}

\section{GHSA Missing Links}
\label{sec:ghsa}

This section explores an empirical study of VFCFinder applied to a set of GHSA security advisories missing VFC links.
As stated in Section~\ref{sec:background}, around 63\% (6,159/9,764) of GHSA advisories do not have VFC links.
We pose the research question: \emph{How does VFCFinder work on real-world data with missing VFC links?}

\subsection{Considered GHSA Advisories}
Section~\ref{sec:ghsa_data_collection} details VFCFinder prerequisites: a source code link, a fixed version, and a prior vulnerable version, which are not universally available across advisories.
Approximately \initialMissingGitHubRepo of VFC-lacking GHSA advisories do not contain a source code link.
As described in the Appendix, we created a simple methodology for obtaining the missing GitHub repository.
We can directly locate project links on GitHub using the GHSA package name (e.g., Source Code/Issues/Homepage).
This method yielded source code links for approximately \foundMissingGitHubRepo GHSA advisories missing VFCs.
However, \stillMissingGitHubRepo GHSA advisories continue to lack source code links.
We have submitted this data to GitHub.

Upon resolving the source code link issue, \potentialRemainMissingGH GHSA advisories still lacked VFC links.
During repository cloning, \missingTagsRepo did not use tags to define the commit window (Section~\ref{sec:model}), and not all advisories possessed fixed versions (\stillNotFixed) or previous vulnerable versions (\missingPreviousTag).
Ultimately, VFCFinder could be used on \missingCleanGHSA advisories.

\subsection{Missing Link Results}
Table~\ref{tbl:ghsa_results} provides the results of VFCFinder applied to GHSAs missing VFCs.
We assessed VFCFinder using a random 10\% (\totalReviewedGHSA) subsample of \missingCleanGHSA advisories.
We manually evaluated the results of VFCFinder.
The manual VFC validation involves understanding the advisory's objectives and reviewing the commit thoroughly after determining the fix's intent.

In total, \totalReviewedGHSA advisories and VFCFinder's output underwent manual review.
VFCFinder's output identified the VFC for \totalGHSAFound advisories.
For \totalGHSAMissed cases, the patch link was found but not among VFCFinder's Top-5.
In \totalGHSACouldNotFind cases, the patch link wasn't found in the reported fixed versions, leading us to assume these 25 advisories may have incorrect fixed version data.
For VFCFinder's Top-N recall calculation, we considered advisories where the VFC was found or wasn't in the Top-5, resulting in \totalGHSAFoundAndMissed advisories.
The Top-N recall results were as follows: \totalGHSATopOne at Top-1, \totalGHSATopTwo at Top-2, \totalGHSATopThree at Top-3, and \totalGHSATopFive at Top-5.

\paragraph{Community Contribution} 
As a valuable contribution to the community, we submitted all found patch links (308) back to GitHub.  
The GHSA database welcomes community enhancements to advisories~\cite{ghsa_review}.
The security team at GitHub independently reviews the suggested updates to determine if the security advisory will be updated.
All 308 patches submitted to GitHub were accepted.  

\begin{table}[t]
\centering
\caption{Results based on \totalReviewedGHSA reviewed GHSA Advsiories missing their VFC. Top-N recall is calculated on the \totalGHSAFoundAndMissed advisories VFCFinder found inside and outside of the Top-5.}
\label{tbl:ghsa_results} \small
    \begin{subtable}{.5\linewidth}  \small
        \centering
          \begin{tabular}{p{0.8in}c} \toprule
            \multicolumn{2}{c}{\textbf{GHSA Breakdown}} \\  \midrule
            Total Reviewed  & \totalReviewedGHSA \\
            Inside Top-5  & \totalGHSAFound \\
            Outside Top-5  & \totalGHSAMissed \\
            Could Not Find & \totalGHSACouldNotFind \\ \bottomrule
          \end{tabular}
      \end{subtable}%
      \begin{subtable}{.6\linewidth} \small
        \centering
          \begin{tabular}{p{0.5in}c} \toprule
            \multicolumn{2}{c}{\textbf{Top-N Recall}} \\ \midrule
              Top-1 & \totalGHSATopOne \\
              Top-2 & \totalGHSATopTwo \\
              Top-3 & \totalGHSATopThree \\
              Top-5 & \totalGHSATopFive \\ \bottomrule
          \end{tabular}
      \end{subtable} 
\end{table}


\vspace{1.0em}
\noindent
\begin{tikzpicture}
    \draw (-10,0) node[draw=black,fill=black!2, thick, rounded corners,inner sep=1ex,text width=0.464\textwidth] {
      {\bf Takeaway: }
      The recall of our empirical study of missing VFCs matches the evaluation in Section~\ref{sec:evaluation}, demonstrating generalizability.
      The GitHub security team reviewed and merged all 308 VFCs into GHSA.
      };
\end{tikzpicture}

\subsection{Ethics and Disclosure}
\label{sec:ethics}
Before submitting VFC links to the GHSA database, we contacted our university's IRB to confirm that doing so is not considered human subject research. 
Analyzing publicly available open-source software projects does not need IRB approval as it does not meet the definition of human subjects research, primarily because the data is public and has always been public.
The public knows these vulnerabilities exist, and we are merely enhancing the quality of the security advisories for the public. 

We initially identified missing source code links and provided the data to the GitHub Security Team. 
Due to the amount of data, the GitHub security team recommended we update through their manual advisory update process.
Doing so allows the team to validate and approve changes to existing security advisories more sufficiently.
Additionally, before submitting VFCs to GitHub, we contacted the security team and confirmed they would want the VFC data.
We also noted we would submit approximately 300 updates, and they agreed it would be manageable for the team.
We limited ourselves to around ten daily updates to avoid overloading the security team at GitHub.

\section{Threats to Validity}
\label{sec:validity}

As in any research, we have threats to the validity of our evaluation and results.
There is potential for noisy labels within our ground truth data. 
While we randomly sampled our non-VFC commits to confirm the absence of VFCs from within the set, some could be within the remainder set.
We place trust in the original stakeholders of the reports who filed them with details surrounding the vulnerability patch, such as using the correct VFC links. 
We note that prior research has reported errors in NVD data~\cite{dong2019towards,nguyen2013reliability}.
Finally, VFCFinder cannot detect the VFC Type for labels it has not seen. 
For example, we had no instances of ``Vulnerable and Outdated Components" in our training data.

\section{Related Work}
\label{sec:relwork}
\paragraph{Vulnerability Fixing Commit Identification}
Zhou and Sharma proposed a stack-based classifier for security issue identification based on bug reports\cite{zhou_automated_2017}.
For instance, Sabetta et al.~\cite{sabetta_practical_2018} uses two linear Support Vector Machine models to classify commits using a Bag-Of-Words representation based on the commit message and git diff.
Sabetta et al. reported an F1 score of 64\% on a single Java dataset.
Wang et al.~\cite{wang_detecting_2019} used a voting algorithm on classifier results to detect security patches.
Wang et al. reported a precision of 66.8\% and recall of 79.6\%, calculating an F1 score of 73\%.
E-SPI~\cite{wu2022enhancing} captures the context of code diffs through contextual AST paths and ensembles with the dependency graph of the commit message. 
The authors report an F1 score of 89.5\% on four projects (Linux, FFmpeg, Wireshark, and QEMU) for identifying VFCs. 
HERMES~\cite{giang_hermes_2022} introduces issue request information with a Support Vector Machine (SVM).
HERMES reported an F1 score of 68\% on a single Java dataset.
dataset~\cite{wang2021patchdb} and reports an F1 score of 90\%.
Zhou et al.~\cite{zhou2021finding} create separate classifiers (including CodeBERT) for commit messages and code changes, subsequently integrating the results through a stacking ensemble technique.
Building on this foundation, Nguyen et al.~\cite{giang_hermes_2022} incorporated commit issues as an additional feature for classification.
Vulcurator~\cite{nguyen2022vulcurator} extended the model using CodeBERT to analyze messages, issues, and code diffs.
Vulcurator reported up to an 87\% on a Python dataset.
SSPCatcher~\cite{sawadogo2022sspcatcher} considered three projects (Linux, OpenSSL, and Wireshark) to evaluate their multi-model SVM approaches to be around F1 scores of 90\%.
Hong et al.~\cite{hong2022xvdb} consider multiple data sources, including issue trackers like Bugzilla, GitHub projects, and Stack Overflow.
TMVDPatch~\cite{zhou2023tmvdpatch} relies on the commit message and the patch to identify VFCs and uses an attention-based BLSTM model. 
TMVDPatch uses the call graph and data flow graph from the patch to represent the semantic and syntactic information of the code diff. TMVDPatch was evaluated on the single C/C++ and reported an F1 score of 90\%.
Midas~\cite{nguyen2023multi} introduced a multi-granularity approach, focusing exclusively on code to identify vulnerability fixes at line, hunk, and file levels.
Zhou et al.~\cite{zhou2023colefunda} introduced CoLeFunDa to identify vulnerability fixes at the function level with an AUC of 80\% only on a Java dataset.
VFFinder~\cite{nguyen2023vffinder} introduced an AST graph-based approach for identifying VFCs based only on code changes.
Evaluating against 507 C/C++ projects, VFFinder reported an F1 score of 69\%. 
Zhou et al.~\cite{zhou2023ccbert} introduced CCBERT, a new transformer-based pre-trained model to represent code changes. 
Within a downstream task of identifying bug-fixing commits, they reported a 91.8\% F1 score on a set of Linux bug-fixing patches using just the code.  
Zuo et al.~\cite{zuo2023commit}, using a transformer-based architecture relying only on the commit message, reported an F1 score of 89.1\% across C/C++ projects with commit patches from NVD.
In parallel, Sun et al.~\cite{sun2023silent} confirmed that Codebert with commit messages and code changes provided the best performance in terms of VFC prediction.

While VFCFinder incorporates identifying a VFC, identifying the VFC does not match it to a security advisory. 
We extend prior work by evaluating CodeBERT across nine different programming languages specifically for identifying vulnerability-fixing commits, whereas existing works have mainly concentrated on C/C++, Python, or Java.
Additionally, we have noted that the overall performance of identifying VFCs is equivalent to ours, ranging in an F1 score of around 90\%.

\paragraph{Vulnerability Fixing Commit Type}
Related to our work has been identifying the type of vulnerability fixed during a commit, but the vast majority has been identifying CWE types for longer descriptions in security advisories~\cite{na2017study, ruohonen2018toward, aota2020automation, das_v2w-bert_2021, pan2023automatic}.
TreeVul~\cite{pan2023fine} uses a CodeBERT to embed the removed and added code during a git diff, which is then fed into a hierarchical Bi-LSTM encoder to predict the CWE type of a VFC.  
TreeVul reported a 72\% weighted F1 score at the depth-3 CWE prediction and up to an 85\% F1 score at the depth-1 CWE prediction on 6,541 commits from 1,560 GitHub OSS projects. 
In addition, CoLeFunDa~\cite{zhou2023colefunda}  can categorize the correct CWE type with an F1 score of 50\% and AUC of 85\%. 
Contrastingly, DAA~\cite{dunlap_2023} took a non-ML approach for VFC identification, which, while capable of producing corresponding CWE types, exhibited limited recall due to reliance on Static Application Security Testing (SAST) tools.
While TreeVul, CoLeFunDa, and DAA are similar to a portion of our work, we predict by the OWASP Top 10 with similar performance.

\section{Conclusion}
\label{sec:conclusion}
The completeness of security advisories is crucial for downstream users, yet about 63\% of GitHub Security Advisories lack their patch link.
This paper presents VFCFinder, a tool designed to perform security advisory to VFC matching.
VFCFinder achieved an accuracy of \vfcFinderTotalTopFiveAcc in identifying the correct VFC within five commits.
Our approach demonstrates that a streamlined pipeline and concise features offer superior generalization over complex systems. 
Applied to GHSA advisories lacking VFCs, VFCFinder found \totalGHSATopFive of the VFCs within the Top-5.
GHSA has accepted and merged over 300 submitted VFCs.

\section*{Acknowledgments}
    This work is supported in part by NSF grants CNS-1946273 and CNS-2207008. Any findings and opinions expressed in this material are those of the authors and do not necessarily reflect the views of the funding agencies.

\bibliographystyle{plain}
\bibliography{main_bib}

\section{Appendix}
\label{sec:appendix}

\subsection{Process for identifying the missing source code links}

Each ecosystem contains an online registry (e.g., PyPI -> \url{https://pypi.org/}). 
Using the package name from a GHSA Advisory, we can do a direct lookup in the respective online registry for the package project links (e.g., Source Code/Issues/Homepage) that point to GitHub.
We first provide an example for PyPI based projects, which was common for majority of the ecosystems.
The only different process was for Maven based projects.

\paragraph{PyPI Example}
\begin{enumerate}
    \item Example GHSA-ID: GHSA-m6xf-fq7q-8743
    \item We extract package name: bleach
    \item We then try to parse the project links on the respective online registry: \url{https://pypi.org/project/bleach/}
    \item We extract the homepage  from the online registry -> \url{https://github.com/mozilla/bleach}
    \item We then return the link that points to a GitHub Repository
\end{enumerate}

\paragraph{Maven Example}
Maven based projects were not so simple.
The following steps were followed to identify Maven source code links:
\begin{enumerate}
    \item First, we search for the project using the following API (\url{https://search.maven.org/solrsearch/select?q={groupId}+AND+a:{artifactId}&rows=10&wt=json})
    \item We extract the package name from the GHSA object: org. springframework.security:spring-security-core
    \item We search using the following API: \url{https://search.maven.org/solrsearch/select?q=org.springframework.security+AND+a:spring-security-core&rows=10&wt=json}
    \item We match based on the groupId and artifactID parsed from the package name.
    \item We pull the latest version of the package from the. Example response:
    \begin{enumerate}
        \item Latest Version: 6.0.1
    \end{enumerate}
    \item We pull the POM file for the latest version using the following API \url{https://search.maven.org/remotecontent?filepath=org/springframework/security/spring-security-core/6.0.1/spring-security-core-6.0.1.pom}
    \item We then search the POM file for the SCM tag that points to a GitHub repository:
    \begin{enumerate}
        \item <connection>scm:git:git://github.com/spring-projects/spring-security.git</connection>
    \end{enumerate}
\end{enumerate}

Our process obtained ~56\% of the missing source code links. 
We provided the appropriate source code to the GitHub security team to pull these links for their security advisories.

\end{document}